# Analysis of Activities and Operations in the Current E-Health Landscape in Tanzania: Focus on Interoperability and Collaboration

Alfred Kajirunga
Computation and Communication Science &Engineering
Nelson Mandela African Institution of Science and Tech
Arusha, Tanzania

Khamisi Kalegele
Computation and Communication Science &Engineering
Nelson Mandela African Institution of Science and Tech
Arusha, Tanzania

*Abstract*— Although the basic application of Information and Communication Technologies (ICT) in the Tanzanian health care systems started years ago, still fragmentation of Information Systems (IS) and limited interoperability remain to be big challenges. In this paper, we present an analysis done on the present health care delivery service, HIS and on some of existing eHealth solutions focusing on interoperability and collaboration. Through interviews, questionnaires and analysis on e-health implementations in relation to interoperability and collaboration we have established that, the lack of standard procedures to guide the lifecycle of eHealth systems across the health sector and poor willingness to collaboration among health stakeholders are key issues which hinders the manifestation of the benefit of ICT use in the health sector of Tanzania. Based on the findings, we provide some recommendations with a view to improve interoperability and collaboration.

I. INTRODUCTION

It is widely accepted that the application of information and communication technologies (ICT) in health has enhanced provision of health services across the world [1], [2], [3]. The World Health Organization defines eHealth as "the cost-effective and secure use of information and communications technologies (ICT) in support of health and health- related fields, including health-care services, health surveillance, health literature , and health education, knowledge and research" [4].Regardless of its importance the adoption of eHealth standards in many African countries is still a challenge [5], [6], [7].Tanzania, like other many African countries, its health care system has been facing almost similar problems [8], [9]. Due to the need of good health care delivery services in the society, these problems cannot be avoided and will require fundamental changes in the current health care arrangements [10]. Tanzanian government through the ministry in charge of health sector and social welfare (MoHSW) has developed its strategic plan called the Health Sector Strategic Plan III to guide priority setting and deployment of resources in the health sector [11]. The already initiated Tanzania National eHealth Strategy (2013 – 2018) of the health care system aims to integrate all fragmented information systems (IS) and offer a complete solution that will benefit all interested parties. To achieve this the issue of eHealth standards, systems interoperability and collaboration between different eHealth stakeholders must be given a serious consideration. Taking into account that it is within achieving systems interoperability, agreement on the data standards to be used must be reached. This results to efficient collaboration among different eHealth stakeholders in accomplishing a number of goals like the improvement of the quality of patient care, reduction of medical errors, and therefore savings in terms both of human and financial costs [5]. A recent study by Lawrence explains the issues, challenges and opportunities towards EHR interoperability in Tanzania hospitals; the main concerns were privacy, security and confidentiality issues when considering information sharing and data sharing [12]. Hence it was important to know how far we are in eHealth standards adoption, systems interoperability and collaboration among eHealth stakeholders in our health sectors.

However, a well formation of the Tanzania health care system should provide opportunities for high quality and professional work with patients and long-term development, whereas relevant and reliable economic, administrative and medical data provided by eHealth should facilitate better quality planning, control and management of individual health care organizations and health care system in general. The focused question answered in this research is: what is currently existing in the Tanzania eHealth landscape? The main objectives of the paper is on analysis of activities and operations in the current eHealth landscape in Tanzania focusing on systems interoperability and collaboration between eHealth stakeholders. After the introduction, the second section of the paper presents the healthcare system of Tanzania's mainland where we see the challenges in adapting eHealth standards in Tanzania. Our study and methodology is in the third section. Fourth section outlines analysis where findings deduced from analysis of activities and operations in the current e-health landscape in Tanzania is presented. Section five provides discussion. The last section is conclusion and recommendation where we provide some recommendations for more effective further development and implementation of eHealth in Tanzania.

## II. HEALTHCARE SYSTEM OF TANZANIA'S MAINLAND

Tanzanian mainland health infrastructure and healthcare services are categorized into four levels; primary level (village health posts, dispensaries, and health centers) to district hospitals, regional hospitals and finally, consultant /specialized hospitals [13]. About 90% of the population live within five kilometers of a primary health facility [10]. The first line care in rural areas is provided by Clinical Officers with 3 years of medical training or Assistant Medical Officers with additional 2 years medical training [14]. The introduction of mandatory health-insurance schemes for formal-sector employees, offering comprehensive health care benefits to their members, the largest being the National Health Insurance Fund covers civil servants, and on the other hand The National Social Security Fund for private formal-sector employees [15].

### A. Challenges in Adapting E-Health in Tanzania

While the integration of ICT and healthcare has brought a lot of potential benefits, there are many challenges which affect its adoption in Tanzania. Different studies show that inadequate ICT infrastructure, unreliable electric power, low ICT budgets, Lack of coordination on ICT matters among ministries, departments, and agencies (MDAs), as well as partners, poor e-healthcare systems design, inadequate ICT skills on the healthcare workers to mention few, are the bottlenecks to the adoption of eHealth in Tanzania [8], [9].As stated in the action plan report by the Ministry of Health and Social Welfare [10]current challenges to eHealth in Tanzania includes:

- A fragmented landscape of eHealth pilot projects and stakeholders
- Numerous data and health information systems (HIS) silos
- Lack of ICT infrastructure
- Lack of ICT workers, in particular those who are well trained
- Lack of coordination on ICT matters among ministries, departments, agencies (MDAs), and the lack of an architecture to guide the development of HIS bottlenecks.
- Lack of compliance with eHealth standards and systems interoperability

With these challenges the analysis of activities and operations in the current e-health landscape in Tanzania was inevitable.

## III. OUR STUDY AND METHODOLOGY

### A. Area of Study

This study was carried out in Dar es Salaam and Arusha, Tanzania. We consider more Dar es Salaam since it has more healthcare facilities as well as key informants from health care workers, preferably supervisors or staff in-charge in health institutions [16].The analysis on HIS was carried out in hospitals, dispensaries(health institutions) and some company that are involved themselves in developing health management systems.

### B. Sampling and Data Collection

A cross-sectional study was deployed in eight hospitals, seven dispensaries and some company that are involved themselves in developing health management systems. Data collection included the use of structured questionnaires and interviews. Data was collected in order to analyze the current activities and operation in eHealth. Guided questionnaires were used to measure the intensity and strength of the factors associated with the current activities and operation in eHealth. Review of existing documents such as journal articles and official reports related to the topic under study was done.

### C. Data Analysis

Statistical Package for Social Sciences (SPSS) was used for data analysis. We present the findings in tables for easy readability and interpretation of data. The significance was tested using a p-value of $p = 0.05$ with a confidence interval of 95%.

## IV. ANALYSIS / RESEARCH FINDINGS

The analysis done on current health care delivery service, applicability of eHealth components and on some of existing eHealth solutions and systems focusing on collaboration and system interoperability, resulted into key findings that are presented in category wise as follows:

### A. Existing eHealth Solution and Health Information Systems

Health service, particularly when considering eHealth (a case of applications and systems) involves several tasks (Reporting, collection, management, knowledge transfer or analysis of data to mention a few). Our examination, reveals the existence of various systems that are concentrating on collection, management and analysis of data, but which are not interconnected and inter-operable.

AllseeEHR system which is implemented in government hospitals in Kinondoni Municipal in Dar es Salaam, it is more about recording of patient information on reporting, but the emphasis is more on recording cash flow from different sectors, although patient history can be viewed once he/she provides his/her registered id but also it is neither inter-operable nor interconnected among the implementing partners. On the other hand, some open source software like OpenMRS and Care2x have been implemented in some areas for various purpose like management of HIV/AIDS, and for registration. LIS, JIVA, LMIS, DHIS2 and CTC2 are present health systems that are implemented in various health institutions but, they are not interoperable or interconnected either.

## B. Distribution of eHealth Services between Rural and Urban Areas

In recent years there has been an increase in the number of health facilities in the country, so that the majority of the population lives within 5 km from a health facility. However, there are still geographical inequalities in access to health services [16].

In relation to geographical inequalities in access to health services (Between rural and urban areas) there is relatively higher support from various stakeholders in urban areas than in the rural areas [8]. This is in line with our findings where a number of health stakeholders prefer to settle their business in urban due to infrastructure problems present in rural areas. This increases the gap we see in access to health services and eHealth applicability between these two areas. As this stands, there is less effort by the government or other stakeholders in health to resolve the situation the challenge being inadequate resources.

## C. Collaboration Among eHealth Stakeholders

Health sector involves a number of stakeholders covering from government, public/user, policy maker, healthcare professionals, Funders etc, who may be categorized differently. In this study, especially when considering collaboration among eHealth stakeholders, we have presented four categories as: developer, implementers, clinicians (Health care provider) and users.

The study revealed that collaboration among the mentioned categories do exist, however the lack of a standardized way (agreed upon tool) for collaboration among the eHealth stakeholders was found to be a big challenge. The result of chi-square test shows that collaboration among eHealth stakeholders level is significant ($p = 0.026$). Also we found out that, there is poor willingness towards collaboration among private companies or vendors who are involved themselves with developing of health management systems (when considering developers). Some of the reasons to this are due to business issues, and there is no initiative so far trying to call those companies together so that they can seat and reach an agreement on how to collaborate, tools for achieving such collaboration, business issues and policy to guide them in their collaboration. This would help to solve the two prior challenges. On that perception we asked the stakeholders (participants) about collaboration and tools used in achieving such collaboration. The results were as follows:

Table 1: Stakeholder's response towards collaboration and tools used in achieving such collaboration (N=102)

| Category | Interviewed stakeholders | Number of collaborating stakeholders | Tools used (Per percentage % representation) | | | | Total % |
|---|---|---|---|---|---|---|---|
| | | | Phone | Email | Phone and Email | Git / any CVS | |
| Developers | 16 | 13 | 43.75 | 25 | ✓ | 12.5 | 81.75 |
| Implementers | 7 | 7 | 71.43 | 28.57 | - | - | 100 |
| Clinicians | 22 | 22 | 54.55 | 45.45 | ✓ | - | 100 |
| Users | 57 | 51 | 80.7 | 8.77 | ✓ | - | 89.47 |

Unreliability of the internet in most of the hospitals regardless of the presence of National ICT Broadband Backbone (NICTBB), results in information exchange by using emails to be less preferred compared to use of mobile phones. Looking into another angle, collaboration among private hospitals or private to government was found poor, that is the willingness of those parts to collaborate is poor. Some argued that they are doing business in which they compete thus it is difficult to collaborate with your competitor; nevertheless we present the view of health stakeholders on how collaboration is in their respective organizations.

| | | Frequency | Percent |
|---|---|---|---|
| Valid | Satisfy | 42 | 41.2 |
| | Poor | 36 | 35.3 |
| | Normal | 24 | 23.5 |
| | Total | 102 | 100.0 |

Table 2: Health stakeholder views on collaboration in their respective organization (N=102).

## D. System Interoperability

As stated in a report by the Ministry of Health and Social Welfare (2013-2018) [10] that: Tanzania's HIS are faced with system interoperability problems. We found out that almost 62.5% of complexity in data integration and hence interoperability were in line with our hypothesis that "Interoperability fails because of lack of coordination at all levels of systems development. A well designed collaboration architecture will facilitate coordination which, in turn, will lead to interoperable systems development". Also, the common use of open source systems which were not specifically designed according to our context and environment (Varying in health culture) being a source of fragmentation and lack of interoperability.

Lack of compulsory governance structure and standards to guide the development of eHealth systems across the health sector (an architecture, Security and Data dictionary) top up to interoperability problem [10].With this remark, we observe different systems with different design and data structure which also add to system interoperability problem. Although the creation of a common data warehouse through integration of the diverse information systems into DHIS2 which deals with more data collection and analysis processes is the current focus, the awareness of interoperability and data standard adoption is still low among the health and ICT workers. As 55.9% of interviews personnel when asked about these two parallel things, their response was poor and the result of chi-square test shows that system interoperability is significant with $p = 0.004$.

## V. DISCUSSION

This study reveals that the current eHealth activities in Tanzania mainland's are still faced with a lot of challenges involving systems interoperability and collaboration among eHealth stakeholders. Although there is an eHealth policy to direct what to be done and how, the situation is quite different in most health centers and hospitals. In most cases the reasons being inadequate ICT infrastructures, inadequate resources, poor ICT skills among health workers and budget limitation. These findings support the findings in previous studies [8], [9].

System interoperability is an important aspect towards achieving good health care service delivery [5]. As that fact stands, in our study, we found out that almost 86.3% of the systems are not capable of sharing information (or not interoperable). Several factors were recognized that are concerned with this situation, the common one being most of the systems are designed as per hospital needs and they differ a lot in their data structure or formats. However, querying multiple data sets with different format requires mediated schema which in turn requires scientists to have knowledge of the query syntax [17] that awareness to most of our health IT stuff is still low. We also found out that security and privacy concerns are associated with most of the organizations not willingly to share their data. This is in line with [12] who said that "Tanzania health consumers should be made comfortable by ensuring that the issue surrounding privacy and security of their health records are clearly addressed before taking any further step towards the implementation of interoperable EHRs for health information exchange". In order to deal with interoperability problems a common data standard must be agreed upon. "At the most basic level, the data standards are about the standardization of data elements: (1) defining what to collect, (2) deciding how to represent what is collected (by designating data types or terminologies), and (3) determining how to encode the data for transmission" [18].So where there is no data standards and data quality, interoperability is becoming a big challenge to handle.

On the other hand, our study reveals that 91.18% of interviewed eHealth stakeholders based on the level of collaboration as defined in this study are capable of collaborating regardless of what tools they are using to achieve such collaboration as shown in Table 1.When rating the existence of collaborating in their respective organization, the results were 41.2% are satisfied, 35.3% rate poor, 23.5% rate normal, respectively. Looking into tools for collaboration, we found out that phones were leading with 73% following with phone and emails 16%, emails 9% and 2% for version control specifically here we considered Git. These findings are consistent with some findings of previous study when giving an account on the adoption and use of ICT by healthcare workers, which report that " Over 93% of the health care institutions use mobile phones in this regard " [8].

Furthermore, we looked into the defined level of collaboration starting with developers from different health organizations (in most cases, they are under ICT department), we found out that they are aware of the existence of other tools like GIT or other versions control systems (VC) for collaboration, but there is no applicable tools so far among them for the purpose of collaboration due to a number of reasons mostly being issues surrounding privacy and security of their health records. This agrees with the study done by Ndume which reports that despite the existence of several collaboration tools naming them as Ning (aimed for network expansion), public library of science (knowledge expansion), Epic surveyors (remote functionality), Scribed (research promotion) as well as Skype, Wiser, Twitter and Facebook, some of the kits don't give researchers peace of mind with respect to security, intact and credibility of their work [17].The situation is the same not only to researchers but also to other different health stakeholders. About 79% of interviewee showed that response, but this is a more traditional way of thinking that can be changed with proper knowledge on those tools and on how to customize them based on their requirements in terms of security and privacy. In the same way we considered the level of clinicians(Health care providers) and users, the interview with them revealed that they have an awareness concerning collaboration even though it is mostly done through mobile phones.

At this point we argued why mobile phones are more involved. The answers were obvious. Any member can buy a phone and found him or herself in one way or another using it as a tool for collaborating with other members in the field. Also, poor or inadequate ICT infrastructures in most of the hospitals resulting in the use of the mobile phone as a number one tool for collaboration. In addition, the "Tanzanian health sector is characterized by a fragmented landscape of ICT pilot projects and numerous data and health information system (HIS) silos with significant barriers to the effective sharing of information between healthcare participants" [10].Hence it is clear that we have the problem of system interoperability and it was observed during the study that in some cases collaboration out of using mobile phones or emails as a tool for achieving collaboration ,it was hindered simply because systems were not capable of sharing information and poor willingness towards collaboration among different health stakeholders due to some reasons revealed in this study. But regardless of several challenges under this area, the need of collaboration and connection of a widespread network of stakeholders within the health care system and between the different health stakeholders level is important, as it was reported in [19] that "Realization of health care sector goals of the vision 2025 needs collaboration of all the key stakeholders involved in health".

This calls for proper technology improvements, especially when dealing with interoperability, collaboration, security and privacy issues, as health data information is highly sensitive and different health organizations have their own orientation, rules and policy. Although agreement on a mechanism for ensuring privacy and security of their health records, technological means and policy to be used may be reached, we must take into account that collaboration is something that cannot be forced but can be agreed upon.

## VI. CONCLUSION AND RECOMMENDATION

In this paper we report about an analysis of current operation and activities in Tanzania mainland's eHealth landscape focusing on interoperability and collaboration. Taking into account that analysis of activities and operation in eHealth landscape is an ongoing activity that needs time and resources, we selected key areas and features in order to meet the objectives and the reality of the situation on the field which was very important in this study. We found that it is important that the introduction of ICT curriculum or ICT training sessions targeting eHealth in health training institutions to health workers has to be considered. By doing this the awareness and effectiveness use of ICT among the health staffs will increase and facilitate its adoption by leveraging the presence of National ICT Broadband Backbone (NICTBB).There is a need for more effort by the government through the ministry in charge of the health sector towards collaboration by promoting this tradition among different health stakeholders. Also, different seminars regarding interoperability issues are to be organized aiming at increasing its IT literacy among health professions. On the other hand, inadequate support, budget limitation, security concerns and unreliable power supply were found to be the most common challenges facing the eHealth activities, a proper attention must be given to these challenges.

AUTHORS PROFILE

Alfred kajirunga, is a Master's student at the Nelson Mandela African Institution of Science and Technology. He is pursuing a master's degree in Communication Science and Technology. He currently lives in Tanzania

Khamisi Kalegele is a lecture at Nelson Mandela African Institution of Science and Technology school of Computational and Communication Sciences and Engineering (CoCSE). He currently lives in Tanzania.